# NIR-to-NIR lifetime based thermometry with the thermally elongated luminescence kinetics driven by structural phase transition in LiYO$_2$:Yb$^{3+}$

M. T. Abbas[1], M. Szymczak[1], V. Kinzhybalo[1], M. Drozd[1], L. Marciniak[1]

[1] Institute of Low Temperature and Structure Research, Polish Academy of Sciences,

Okólna 2, 50-422 Wrocław, Poland

**Abstract**

Among the various techniques used in luminescence thermometry, luminescence kinetics is considered the least sensitive to perturbations related to the optical properties of the medium containing the phosphor. For this reason, temperature sensing and imaging using lifetime-based luminescence thermometers is of high interest for wide range of specific applications. However, for most such thermometers, an increase in temperature leads to a shortening in lifetime, which can hinder the specificity and accuracy of the readout. In this work, we present an approach that utilizes a thermally induced increase in the symmetry of the host material associated with a structural phase transition in LiYO$_2$:Yb$^{3+}$. Consequently, the lifetime of the excited level $^2$F$_{5/2}$ of the Yb$^{3+}$ ion is thermally prolonged, achieving a relative sensitivity of 0.5%/K. The phase transition temperature can be controlled by adjusting the dopant concentration. Additionally, thermal changes in the emission spectrum enable the use of LiYO$_2$:Yb$^{3+}$ for ratiometric temperature readout with a relative sensitivity of 5.3%/K at 280K for LiYO$_2$:5%Yb$^{3+}$.

**Introduction**

Intensive research on luminescence thermometry conducted over the past decade has established it as the foremost technique for remote temperature determination[1–5]. Although many different spectroscopic parameters exhibit thermal variability, practical and reliable temperature determination is typically achieved using only two approaches: the ratiometric and lifetime-based thermometry[6–8]. Both techniques have experienced significant development in recent years, with various mechanisms employed to enhance the thermal dynamics of observed changes and, consequently, increase the sensitivity of luminescence thermometers[9,10]. The most commonly used mechanisms in temperature sensing include thermalization of energy states[10–13], phonon-assisted energy transfer between dopant ions[14–17], thermal activation of additional quenching channels[18], and thermal sensitization[19,20]. One particularly interesting and recently proposed solution is luminescence thermometry based on inorganic host materials that exhibit thermally induced first-order phase transitions[21–27]. In this approach, a phase transition changes the point symmetry of the site occupied by the luminescent dopant, thereby altering the spectroscopic properties of the dopant. Typically, these changes involve variations in the number of Stark components into which the dopant ion's energy levels are split[21–27]. As a result, an increase in temperature leads to the successive quenching of luminescence associated with the low-temperature phase and an enhancement in the intensity of emission bands originating from the high-temperature phase. The ratio of their intensities can be effectively used for ratiometric temperature measurement, with the dynamic nature of these changes enabling the achievement of very high relative sensitivities (exceeding 10%/K)[22,23,27]. However, a limitation of this approach is the narrow operating thermal range, as the described structural changes are observed only near the phase transition temperature[25,26]. Recent studies have demonstrated that by appropriately modifying the material's structure to adjust the phase transition temperature, it is possible to develop a comprehensive library of phase-transition-based luminescence thermometers with operating ranges spanning a wide thermal range[25,26]. While this solution is highly attractive, the ratiometric approach has limited applicability under certain conditions[7]. Specifically, when the phosphor is in a medium with strong light absorption or scattering, the medium's dispersion dependence of the extinction coefficient can significantly alter the emission spectrum shape, thus affecting the reliability of the temperature readout. In such cases, lifetime-based luminescence thermometers are preferred, as these conditions do not significantly impact reading reliability[6,7,28].

Although the ratiometric approach for phase-transition-based luminescence thermometers has been described in several papers, to our knowledge, there are no reports on thermally induced changes in lifetime-based thermometers using materials characterized by a first-order structural phase transition. It is expected that a change in the point symmetry of a luminescent ion can significantly affect the probability of radiative depopulation of the excited level and thus the kinetics of its luminescence. For example, in a material like $LiYO_2$, an increase in temperature results in a symmetry change from the monoclinic structure observed at low temperatures to a tetragonal structure at high temperatures (Figure 1)[29–32]. Consequently, the point symmetry increases, which should lead to an elongation of the lifetime of the emitting level of dopant ion.

In this work, we present a lifetime-based luminescence thermometer based on the kinetics of $^2F_{5/2}$ level depopulation of $Yb^{3+}$ ions in $LiYO_2$ (Figure 1). The described increase in the point symmetry of the $Yb^{3+}$ ion from $C_2$ to $D_{2d}$ reduces the probability of radiative depopulation of the emitting level, thereby inducing prolongation of the luminescence decay profiles. This is unique and, unlike typical lifetime-based luminescence thermometers, represents the first instance where an increase in temperature results in elongation in lifetime[33–36]. Significantly, this proposed strategy allows modulation of the operating thermal range of such a thermometer by changing the phase transition temperature through varying the concentration of $Yb^{3+}$ ions in doped $LiYO_2$.

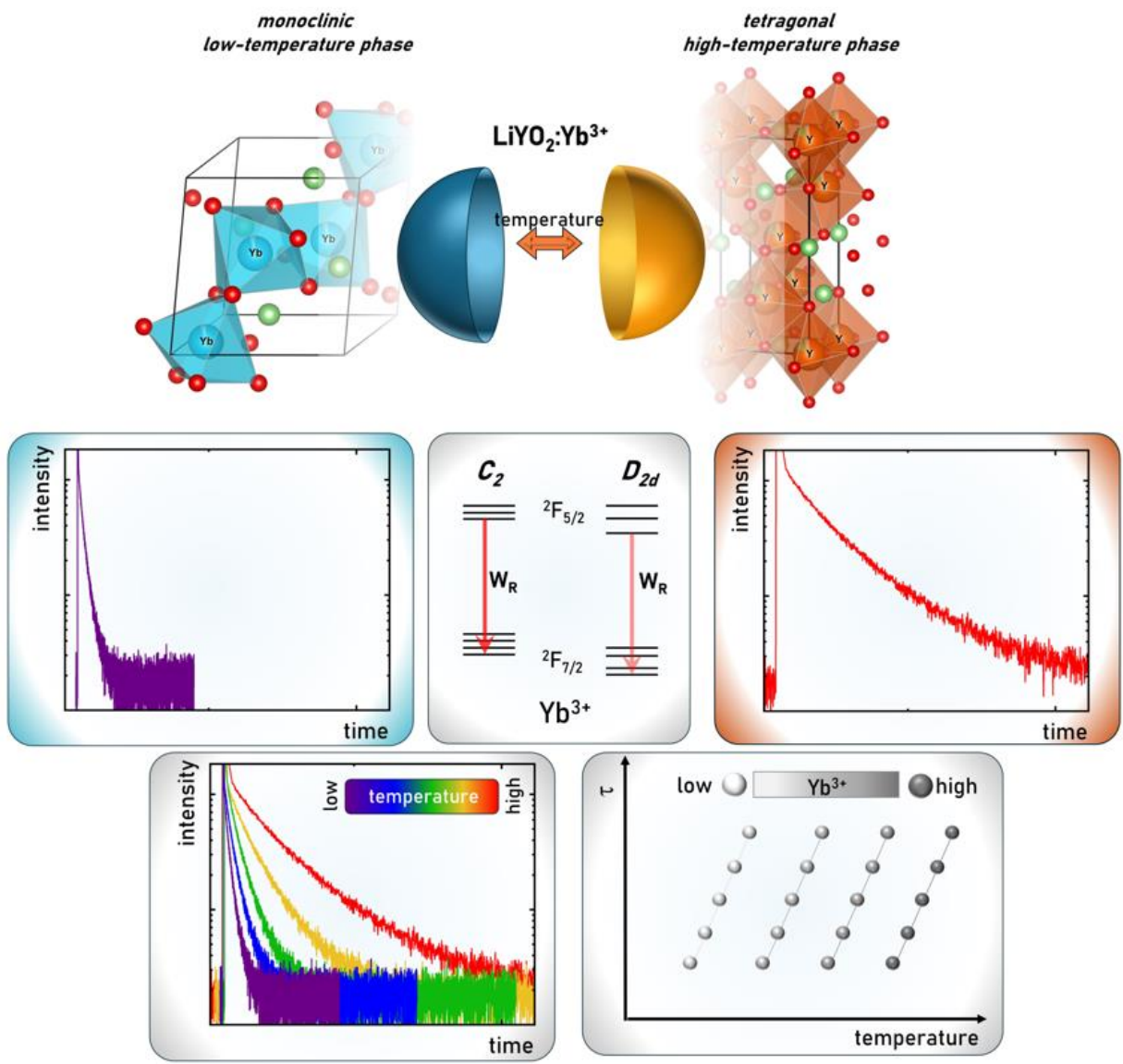

**Figure 1**. The conceptual illustration of our proposed strategy of luminescence lifetime thermometry based on structural phase transition.

**Experimental section**

*Synthesis*

The powders of nanocrystalline LiYO$_2$:x%Yb$^{3+}$, where x = 0, 1, 2, 5, 10, 20, were synthesized using a modified Pechini method[37]. Li$_2$CO$_3$ (99.9% of purity, Chempur), Y$_2$O$_3$ (99.999% of purity, Stanford Materials Corporation), Yb$_2$O$_3$ (99.99% of purity, Stanford Materials Corporation), C$_6$H$_8$O$_7$ (CA, >99.5% of purity, Alfa Aesar) and H(OCH$_2$CH$_2$)$_n$OH, (PEG-200, n = 200, Alfa Aesar) were used as starting materials without any additional purification. Lanthanide oxides were dissolved in deionized water with the addition of 3 cm$^3$ of HNO$_3$ (65% solution, Avantor), then recrystallized three times to remove the excess of nitric acid. After recrystallization, the 4-fold stoichiometric excess of lithium carbonate was added to the water solution of lanthanide nitrates. After that, citric acid (6-fold molar excess in relation to metal cations) and PEG-200 (a molar ratio of 1:1 in relation to CA) were added. Subsequently, the obtained solutions were dried for 3 days at 363 K until a resin was formed. The resins thus prepared were annealed in porcelain crucibles for 6 h in air at 1273 K (heating rate of 10 K/min). The resulting white powders were thoroughly ground in an agate mortar and left for further analysis.

*Characterization*

The obtained materials were examined using powder X-ray diffraction technique. Powder diffraction data were obtained in Bragg–Brentano geometry using a PANalytical X'Pert Pro diffractometer equipped with Oxford Cryosystems Phenix (low-temperature measurements) attachment using Ni-filtered Cu K$\alpha$ radiation (V = 40 kV, I = 30 mA). The sample for variable-temperature X-ray diffraction measurements was mixed with Apiezon grease. Diffraction patterns in 2θ range of 15-90⁰ were measured in cooling/heating sequence in the temperature range from 320 to 80 K. ICSD database entries No. 50,992 (LT phase) and 50,993 (HT phase) were taken as initial models for the analysis of the obtained diffraction data. The morphology of obtained crystals was analysed with Philips CM-20 SuperTwin transmission electron microscope (TEM), operating at 160 kV. The sample was ground in an agate mortar and dispersed in methanol. A drop of the prepared suspension was placed on a copper microscope grid covered with carbon. Before the measurement, the sample was dried and purified in a $H_2/O_2$ plasma cleaner for 1 min. A differential scanning calorimetric (DSC) measurements were performed using Perkin-Elmer DSC 8000 calorimeter equipped with Controlled Liquid Nitrogen Accessory LN2 with a heating/cooling rate of 10 K/min. The sample was sealed in the aluminum pan. The measurement was performed for the powder sample in the X – X K temperature range. The excitation spectra were obtained using the FLS1000 Fluorescence Spectrometer from Edinburgh Instruments equipped with 450 W Xenon lamp and R928 photomultiplier tube from Hamamatsu. Emission spectra were measured using the same system. During the temperature-dependent emission measurements, the temperature of the sample was controlled by a THMS600 heating–cooling stage from Linkam (0.1 ⁰C temperature stability and 0.1 ⁰C set point resolution). Luminescence decay profiles were also recorded using the FLS1000 equipped with 150 W μFlash lamp. The average lifetime ($\tau_{avr}$) of the excited levels was calculated based on fit of the luminescence decay profiles by double-exponential function:

$$\tau_{avr} = \frac{A_1 \tau_1^2 + A_2 \tau_2^2}{A_1 \tau_1 + A_2 \tau_2} \qquad (1)$$

$$I(t) = I_0 + A_1 \cdot \exp(-\frac{t}{t_1}) + A_2 \cdot \exp(-\frac{t}{t_2}) \qquad (2)$$

where $\tau_1$ and $\tau_2$ represent the luminescence decay parameters and $A_1$, $A_2$ are the fitted amplitudes of the double-exponential function.

**Results and discussion**

The LiYO$_2$ belongs to the family of compounds of a general formula Li$A$O$_2$ (where $A$ represents the rare earth ions)[29–32]. The members of this family of compounds crystalize in the three different structure types: (1) tetragonal of $I4_1/amd$ space group (Z = 4), (2) monoclinic $P2_1/c$ and (3) orthorhombic of $Pbnm$ space group (Z = 4)[38]. However in the case of LiYO$_2$ only two crystallographic phases are observed: low-temperature monoclinic structure of $a$ = 6.1493(8) Å, $b$ = 6.1500(10) Å, $c$ = 6.2494(2) Å and $\beta$= 119.091(5)° and high-temperature tetragonal structure of $a$ = 4.4468(9) Å, $c$ = 10.372(22) Å cell parameters (Figure 2a). In the case of the bulk crystals the phase transition was reported at 363 K[38]. However its temperature can be modified by the dopant ions and their concentration (depending on the difference in ionic radii between dopant and host material ions) and the sample morphology (decrease of the temperature of the phase transition with the reduction of the particle size). Due to the similarity in ionic charge, radius, and preferred coordination, Yb$^{3+}$ ions substitute Y$^{3+}$ ions in LiYO$_2$:Yb$^{3+}$ and the phase transition modifies the point symmetry from $C_2$ to $D_{2d}$ with temperature increase[39]. Comparison of XRD patterns obtained for LiYO$_2$ with different concentrations of Yb$^{3+}$ ions with reference patterns allows to conclude that all obtained materials are phase-pure, with no additional reflections observed (Figure 2b). XRD patterns measured at room temperature unambiguously correspond to the high-temperature tetragonal phase of LiYO$_2$. As the concentration of Yb$^{3+}$ increases, a slight shift of the reflections towards larger 2θ angles is observed, which is related to the successive reduction of the unit cell parameters due to the difference in ionic radii between the host material ion $Y^{3+}$ (R=0.900 Å) and the $Yb^{3+}$ dopant ion (R=0.868 Å)[40].

To investigate the rate of structural phase transition between low and high-temperature phases of $LiYO_2$, XRD measurements were performed for phosphors with extreme $Yb^{3+}$ ion concentrations of 1% and 20% as a function of temperature (Supporting Information). Rietveld analysis of the obtained patterns allowed to determine the ratio of high-temperature (HT) to low-temperature (LT) phases as a function of temperature (Figure 2c). Regardless of the $Yb^{3+}$ concentration, an increase in temperature results in a successive increase in HT phase content, and above 230 K (for 20% $Yb^{3+}$) and 320 K (for 1% $Yb^{3+}$), only the high-temperature phase is observed in the analyzed XRD patterns. The contribution of high-temperature phase begins to dominate (HT/LT > 50%) at approximately 225 K and 290 K for 20% $Yb^{3+}$ and 1% $Yb^{3+}$, respectively (Figure 2d). The pronounced effect of dopant ion concentration on the phase transition temperature is an expected phenomenon, described for many materials, and is related to the aforementioned difference in ionic radii between host material and dopant ions[22,23,27]. These values correspond to the phase transition temperatures for examined materials obtained from the DSC study (Figure 2e). Additionally, the typical for first-order phase transition hysteresis within heating and cooling cycle were observed. For the undoped system, the phase transition occurs at 293.5 K during heating and 283.9 K during cooling, while for the $LiYO_2$:20% $Yb^{3+}$ sample, it occurs at 223 K during heating and 217.7 K during cooling. Analysis of the morphology of the obtained materials indicates that the concentration of $Yb^{3+}$ ions has no significant effect on the average size and its distribution. The obtained $LiYO_2$:$Yb^{3+}$ consists of aggregated nanoparticles with an average size of 52 ± 7 nm (Figure 2f-i).

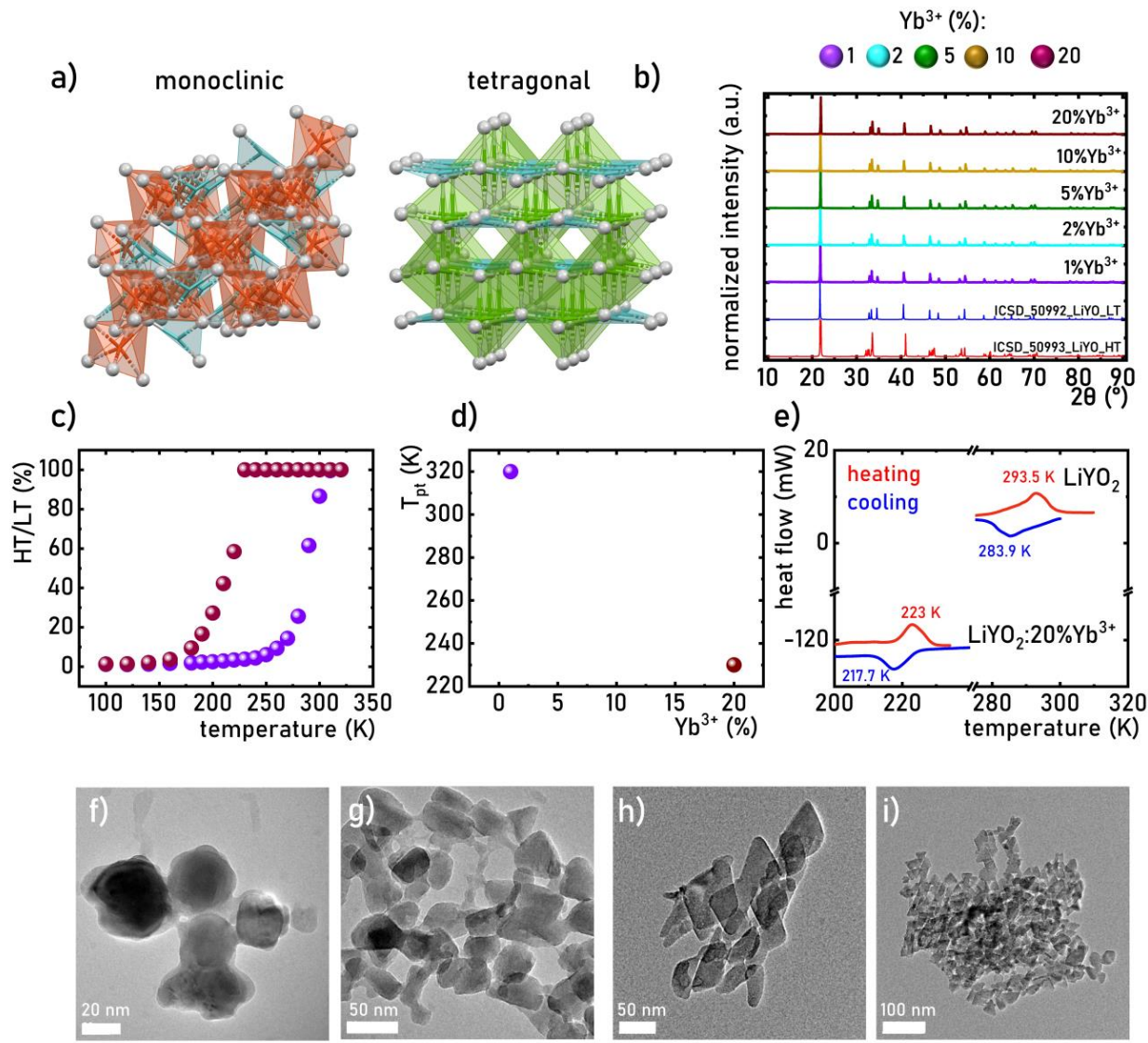

**Figure 2.** Visualization of the structure of low-temperature (LT) monoclinic and high-temperature (HT) tetragonal phases of $LiYO_2$-a) the comparison of the room temperature XRD patterns for $LiYO_2$:$Yb^{3+}$ with different $Yb^{3+}$ concentration-b); the ratio of the HT to LT phases in the $LiYO_2$:$Yb^{3+}$ determined from the XRD patterns as a function of temperature-c); phase transition temperature (determined from XRD data) as a function of $Yb^{3+}$ concentration-d); the comparison of the DSC curves for $LiYO_2$ and $LiYO_2$:20%$Yb^{3+}$-e); representative TEM images for the $LiYO_2$:1%$Yb^{3+}$-f), g) and $LiYO_2$:20%$Yb^{3+}$- h), i).

The energy diagram of the $Yb^{3+}$ ion is notably limited, consisting of only one excited level, $^2F_{5/2}$, and one ground level, $^2F_{7/2}$ (Figure 3a)[41–44]. Consequently, the $^2F_{5/2} \rightarrow {}^2F_{7/2}$ electronic transition is characterized by high oscillator strength[41]. Due to the interaction of the $^2F_{5/2}$ and $^2F_{7/2}$ multiplets with the crystal field of the host material, they split into four (marked as 1-4) and three (5-7) Stark components, respectively. Although the number of these components does not depend on the point symmetry of the ion, the strength of the splitting and the energies of the Stark levels can be affected by the host material. Typically, an increase in point symmetry results in increased splitting energy, causing a spectral shift of the emission lines (Figure 3b). Analysis of the emission spectra for $LiYO_2:Yb^{3+}$ in both low-temperature and high-temperature phases confirms this behavior. In addition to the obvious thermal broadening of the Stark lines, slight shifts in the energy of individual lines are observed. Deconvolution of the spectra revealed eight components, suggesting that emission transitions occur not only from component 5 but also from level 6 of the $^2F_{5/2}$ to all Stark components of the $^2F_{5/2}$ level (Figure 3c). This large number of components is uncommon for the emission spectra of $Yb^{3+}$-doped phosphors but has been previously observed in nanocrystalline $LaPO_4:Yb^{3+}$ [45], likely due to thermalization of level 6, resulting from the small size of the phosphor impeding light-induced heat dissipation.

Detailed analysis of the emission and excitation spectra enabled the generation of $Yb^{3+}$ energy diagrams for the monoclinic and tetragonal phases of $LiYO_2$. An increase in $Yb^{3+}$ concentration does not significantly alter the shape of the emission spectra measured at room temperature (293 K) (Figure 3d). These spectra reflect the spectroscopic properties of the high-temperature phases, consistent with structural studies. However, for $LiYO_2:20\%Yb^{3+}$, a notable reduction in the emission intensity of the lines 6→1 (970 nm) and 5→1 (975 nm) in respect to the emission at 991 nm can be observed, which may be due to the reabsorption process. As $Yb^{3+}$ concentration increases, the average interionic distance decreases, facilitating reabsorption, particularly affecting resonance transitions.

Luminescence kinetics analysis reveals that at low $Yb^{3+}$ concentrations (below 5%), the luminescence decay profiles exhibit a single exponential shape (Figure 3e). However, at concentrations above 10%, the decay profiles deviate from this trend, and lifetimes are significantly shortened. For a reliable comparative analysis under consistent conditions, average lifetimes ($\tau_{avr}$) were determined as described in the Experimental section (Eq. 1 and 2). The $\tau_{avr}$ decreases from $\tau_{avr} = 0.82$ ms for 1% $Yb^{3+}$ to $\tau_{avr} = 0.22$ ms for 20% $Yb^{3+}$ (Figure 3f). Given the limited energy diagram of $Yb^{3+}$ ions, the observed shortening of $\tau_{avr}$ may occur due to two main reasons: (1) energy transfer to an unintended dopant, with $Er^{3+}$ being the most likely candidate, or (2) energy diffusion across the excited levels of $Yb^{3+}$ to structural or surface defects. Since no up-conversion emission of $Er^{3+}$ in the visible spectrum or characteristic Stokes emission around 1550 nm (associated with the $^4I_{13/2} \to {}^4I_{15/2}$ transition of $Er^{3+}$ ions) was observed for $LiYO_2:Yb^{3+}$, the first hypothesis was excluded. Therefore, the observed shortening of $\tau_{avr}$ is most likely due to energy diffusion, a mechanism often responsible for the quenching of $Yb^{3+}$ luminescence in nanocrystalline materials[46–48]. Furthermore, the short $Yb^{3+}$-$Yb^{3+}$ distance allowing energy reabsorption and altering the emission spectrum shape strongly suggests that energy diffusion is highly probable.

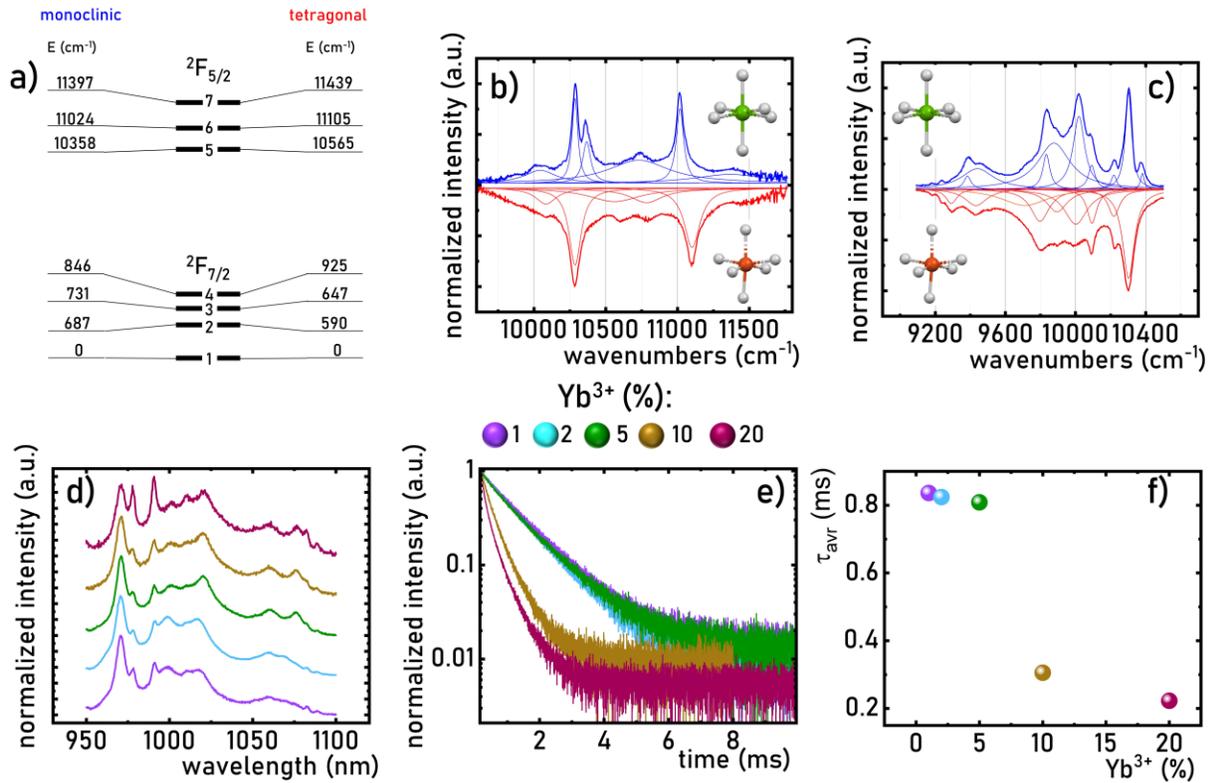

**Figure 3**. Energy levels diagram of $Yb^{3+}$ ions in monoclinic and tetragonal $LiYO_2$:$Yb^{3+}$ phases -a); the comparison of the excitation-b) and emission spectra-c) of $LiYO_2$:1%$Yb^{3+}$ measured at 100 K (blue line) and 333 K (red line); the comparison of the room temperature emission spectra -d) and luminescence decay profiles- e) of $LiYO_2$:$Yb^{3+}$ for different concentrations of $Yb^{3+}$ ions; the influence of the $Yb^{3+}$ concentration on the room-temperature $\tau_{avr}$-f).

Changes in the energy of the Stark levels of both the $^2F_{5/2}$ and $^2F_{7/2}$ levels induce alterations in the shape of the $LiYO_2$:$Yb^{3+}$ emission spectrum as the temperature varies (Figure 4a, Figure S1-S5). To further illustrate these changes for a representative sample of $LiYO_2$:1%$Yb^{3+}$, the normalized emission spectra are presented as a map in Figure 4b. A clear spectral shift of the bands above 310 K can be observed. However, an increase in the emission intensity of lines associated with the high-temperature phase of $LiYO_2$:$Yb^{3+}$ is already noticeable above 200 K. Therefore, the ratio of the emission intensity of the spectral lines of the high-temperature phase to the low-temperature phase of LiYO$_2$:Yb$^{3+}$ (LIR), defined as follows:

$$LIR = \frac{Yb^{3+}(tetragonal)}{Yb^{3+}(monoclinic)} = \frac{\int_{1075nm}^{1077nm} {}^2F_{5/2} \to {}^2F_{7/2} d\lambda}{\int_{996nm}^{998nm} {}^2F_{5/2} \to {}^2F_{7/2} d\lambda} \quad (3)$$

can be used for ratiometric temperature readout. Analysis of the thermal dependence of the LIR (data presented in Figure 4c, normalized to the LIR value at 77 K) shows that, regardless of Yb$^{3+}$ concentration, a sharp increase in the LIR value is observed when the threshold temperature is exceeded, followed by a slight decrease when the maximum value is reached. Since a reliable temperature reading can only be obtained under conditions of monotonic change of the thermometric parameter, such thermal behavior of LIR narrows the operating temperature range.

An increase in Yb$^{3+}$ concentration causes a gradual decrease in the temperature at which changes in LIR values are observed. This effect is due to the difference in ionic radii between Y$^{3+}$ and the dopant ion Yb$^{3+}$, and the associated change in phase transition temperature previously described. Treating the temperature at which the LIR reached its maximum value as the phase transition temperature ($T_{pt}$), one can observe a linear decrease from 302 K for 1% Yb$^{3+}$ to 208 K for 20% Yb$^{3+}$ (Figure 4d). This directly demonstrates how the thermometric performance of this ratiometric luminescence thermometer based on LiYO$_2$:Yb$^{3+}$ can be modulated by changing the dopant ion concentration. The rate of thermal LIR changes affects the thermometric performance of the luminescence thermometer. To quantify the observed changes, the relative sensitivity ($S_R$) is determined:

$$S_R = \frac{1}{LIR}\frac{\Delta LIR}{\Delta T}\cdot 100\% \qquad (4)$$

where ΔLIR represents the change in LIR corresponding to the change in temperature by ΔT. The maximum relative sensitivity was obtained for 5% $Yb^{3+}$, with a value of 5.3%/K (Figure 4e). No regular change in $S_R$ values was registered with varying $Yb^{3+}$ ion concentration. However, for each concentration analyzed, $S_{Rmax}$ exceeded 3.8%/K. The narrow thermal range over which an increase in LIR values was observed is reflected in the narrow operating thermal range for which positive $S_R$ values are recorded, varying from 200-310 K for 1% $Yb^{3+}$ to 150-230 K for 20% $Yb^{3+}$.

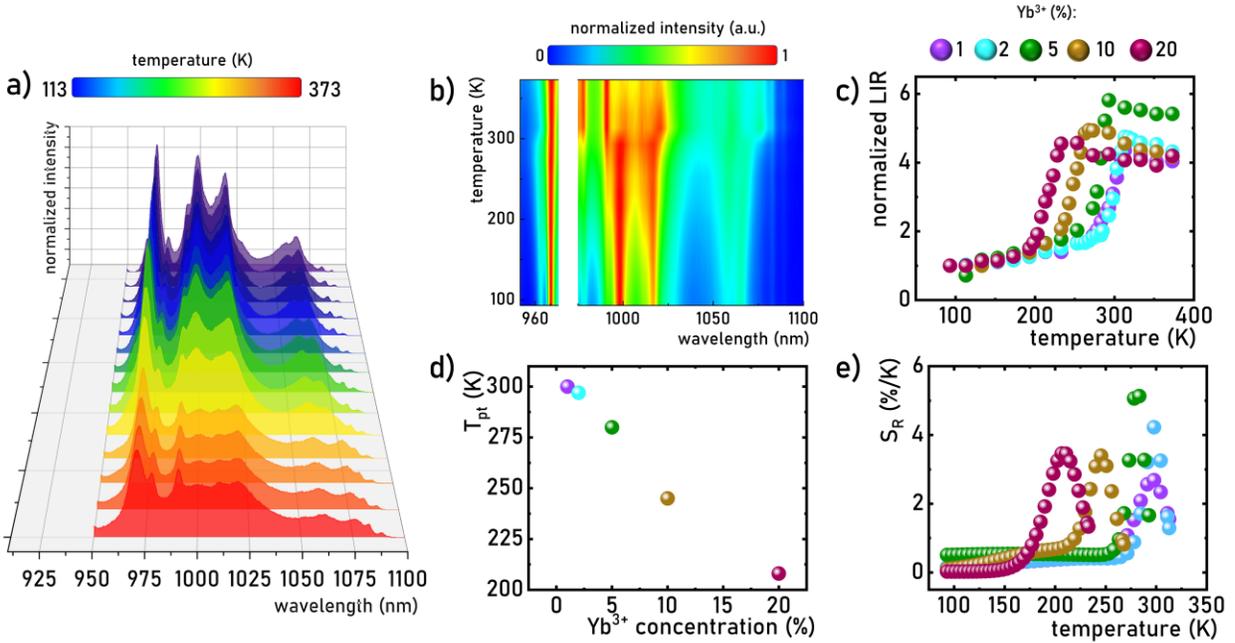

**Figure 4.** Luminescence spectra of $LiYO_2$:1% $Yb^{3+}$ measured as a function of temperature -a); and luminescence map of normalized emission spectra -b); the thermal dependence of normalized LIR for different $Yb^{3+}$ concentrations-c) the influence of the $Yb^{3+}$ on the $T_{pt}$-d); thermal dependence of $S_R$ for the ratiometric luminescence thermometer for different $Yb^{3+}$ concentrations-e).

The structural phase transition from monoclinic to tetragonal in LiYO$_2$:Yb$^{3+}$ changes the point symmetry of the crystallographic site occupied by Yb$^{3+}$. This increase in symmetry can modify the radiative depopulation rate ($W_R$) of the emitting state, thereby affecting the luminescence decay profile. Consequently, the luminescence decay profile of the $^2F_{5/2}$ state of Yb$^{3+}$ successively prolongates with increasing temperature (Figure 5a, Figure S6-S10). This is an extremely rare thermal effect and stands in contrast to what is observed for most lifetime-based luminescence thermometers, where an increase in temperature typically results in shortening of lifetime due to thermal depopulation of the excited level via multi-phonon processes or phonon-assisted energy transfers[33–36]. To the best of our knowledge, this study is the first report on the use of such an effect for optical thermometry. This effect is observed for each of the analyzed concentrations of Yb$^{3+}$ ions, and the thermal dynamics of the observed changes are similar. Initially, as the temperature increases, rapid changes are not observed. A sharp elongation of $\tau_{avr}$ is observed around the phase transition temperature (Figure 5b). Further temperature increases lead to shortening again, within the thermal range where a decrease in luminescence intensity is also observed. Starting from the basic equation describing the kinetics of depopulation of the excited level:

$$\frac{1}{\tau_{avr}} = W_R + W_{NR} = \frac{1}{\tau_0} + W_{NR} \qquad (5)$$

where $W_{NR}$ and $\tau_0$ represents probability of nonradiative depopulation of the excited state and radiative lifetime of the excited state, respectively. Considering that the energy level scheme of the Yb$^{3+}$ ion consists of only two levels separated by about 10,000 cm$^{-1}$, it can be assumed that the nonradiative decay rate ($W_{NR}$) does not change significantly up to 320 K (Figure 5c). Additionally, given the similar values of $\tau_{avr}$ for 1% and 2% Yb$^{3+}$, it can be assumed that $\tau_0$ (at 77 K) for LiYO$_2$:1%Yb$^{3+}$ is equal to $\tau_{avr}$. Based on this assumptions, the thermal dependence of the radiative decay rate ($W_R$) for LiYO$_2$:1%Yb$^{3+}$ can be determined. It is evident that as the temperature increases up to 260 K, the $W_R$ value remains independent of temperature. Above this temperature, it decreases by about 15% and then stabilizes. This analysis suggests that the phase transition-driven change in $W_R$ is responsible for the observed elongation of the $^2F_{5/2}$ level kinetics of Yb$^{3+}$ ions in LiYO$_2$:Yb$^{3+}$. The described elongation of $\tau_{avr}$ is observed for all Yb$^{3+}$ concentrations, and as with the ratiometric approach, the thermal range over which this effect is observable decreases with increasing Yb$^{3+}$ concentration (Figure 5d). However, as often observed in materials exhibiting a first-order structural phase transition, the inertia of the system within heating and cooling cycles generates a hysteresis loop in the properties of the material[25][24,26]. This is also true for $\tau_{avr}$, where a relatively narrow hysteresis loop is observed (Figure 5e). Such an effect can adversely affect thermometric performance, but the effect is much weaker than that observed for the ratiometric approach in LiYO$_2$:Eu$^{3+}$ [49] or LiYO$_2$:Nd$^{3+}$[50]. The temperatures for which the maximum values of relative sensitivity are observed shift similarly to the ratiometric approach with increasing Yb$^{3+}$ concentration, ranging from 0.6%/K at 290 K for LiYO$_2$:5%Yb$^{3+}$ to 0.32%/K at 150 K for LiYO$_2$:20%Yb$^{3+}$ (Figure 5f).

The sensitivity values obtained for LiYO$_2$:Yb$^{3+}$ are not very high compared to other NIR-to-NIR lifetime-based luminescence thermometry approaches[16,51–55], but this is the first report describing thermal extension of luminescence kinetics. This is important because for typical luminescence thermometers, both an increase in temperature and the presence of quenching centers lead to a shortening of luminescence kinetics, which can hinder reliable measurements under certain conditions.

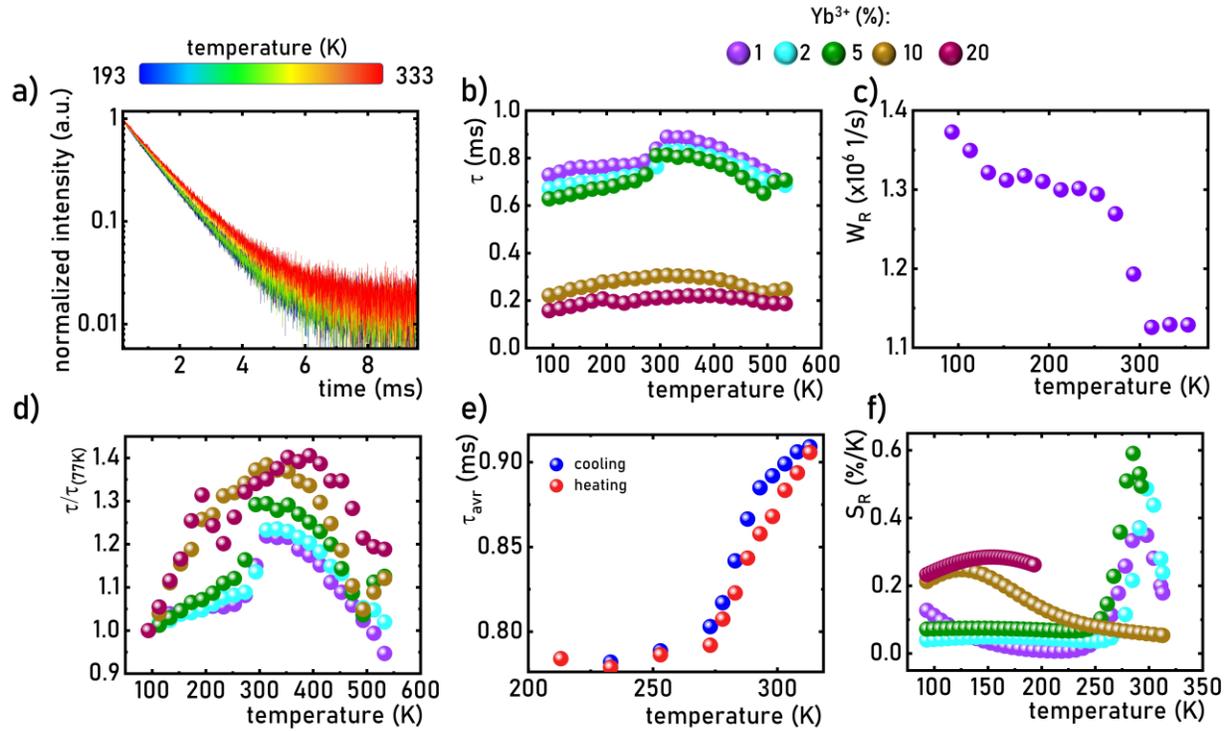

**Figure 5.** Luminescence decay profiles of the LiYO$_2$:1%Yb$^{3+}$ ($\lambda_{exc}$=940 nm, $\lambda_{em}$=980 nm) measured as a function of temperature -a); thermal dependence of the $\tau_{avr}$ for different Yb$^{3+}$ ions concentrations -b) thermal dependence of $W_R$ for LiYO$_2$:1%Yb$^{3+}$ -c); $\tau_{avr}/\tau_{(77\ K)}$ for different concentrations of Yb$^{3+}$-d); the $\tau_{avr}$ of LiYO$_2$:1%Yb$^{3+}$ measured as a function of temperature withing heating and cooling cycle-e); thermal dependence of $S_R$ for lifetime-based luminescence thermometers for LiYO$_2$:Yb$^{3+}$ with different Yb$^{3+}$ concentrations.

**Conclusions**

In this work, the spectroscopic properties of LiYO$_2$:Yb$^{3+}$ were studied as a function of temperature for different concentrations of Yb$^{3+}$ ions. It was shown that the thermally induced structural phase transition from the low-temperature monoclinic to the high-temperature tetragonal phase in LiYO$_2$ resulted in changes in the emission spectrum and the luminescence decay profiles of Yb$^{3+}$ emission. The change in the energies of Stark lines of the $^2F_{5/2}$ and $^2F_{7/2}$ states induced by the phase transition led to alterations in the emission spectra of Yb$^{3+}$ ions. The abrupt thermal change in the ratio of the luminescence intensity of the lines associated with the high-temperature (HT) phase in respect to the low-temperature (LT) phases enabled the developement of a ratiometric luminescence thermometer with a maximum relative sensitivity of 5.3%/K for LiYO$_2$:5% Yb$^{3+}$. An increase in the concentration of Yb$^{3+}$ ions, due to the difference in ionic radii between Y$^{3+}$ and Yb$^{3+}$, resulted in a shift in the phase transition temperature, thereby shifting the thermal operating range of the luminescence thermometers. The thermally induced increase in the point symmetry of the crystallographic site occupied by Yb$^{3+}$ ions from $C_2$ to $D_{2d}$ results in a reduction of the rate of radiative depopulation of the $^2F_{5/2}$ state, thereby elongating the luminescence decay profile corresponding to the $^2F_{5/2} \rightarrow ^2F_{7/2}$ electronic transition of Yb$^{3+}$. The thermal range of the observed elongation shifted towards lower temperatures as the Yb$^{3+}$ concentration increased. The maximum sensitivity of 0.5%/K for the lifetime-based approach was observed for LiYO$_2$:5%Yb$^{3+}$. As demonstrated, the operating thermal range of the LiYO$_2$:Yb$^{3+}$ ratiometric and lifetime-based luminescent thermometers can be adjusted according to the requirements of the application by varying the Yb$^{3+}$ concentration. To the best of our knowledge, this is the first report presenting an optical thermometer based on the thermal elongation of the luminescence decay profile.

**Acknowledgements**

This work was supported by the National Science Center (NCN) Poland under project no. DEC-UMO-2022/45/B/ST5/01629. .